\documentclass[a4paper,fleqn]{cas-dc}

\usepackage[authoryear,longnamesfirst]{natbib}
\usepackage{lineno,hyperref}
\usepackage{ulem}
\usepackage{algorithm2e,setspace}
\usepackage{subfigure}
\usepackage{multirow}
\usepackage{bbding}
\usepackage{amssymb}

\def\tsc#1{\csdef{#1}{\textsc{\lowercase{#1}}\xspace}}
\tsc{WGM}
\tsc{QE}
\tsc{EP}
\tsc{PMS}
\tsc{BEC}
\tsc{DE}

\begin{document}
\let\WriteBookmarks\relax
\def\floatpagepagefraction{1}
\def\textpagefraction{.001}

\shorttitle{Underwater Acoustic Target Recognition based on Smoothness-inducing Regularization and Spectrogram-based Data Augmentation}

\shortauthors{Xu J, Yuan X.}

\title [mode = title]{Underwater Acoustic Target Recognition based on Smoothness-inducing Regularization and Spectrogram-based Data Augmentation}                      




%
\author[address1,address2]{Ji Xu}[
                        style=chinese,
                        orcid=0000-0002-3754-228X]


\ead{xuji@hccl.ioa.cn.cn}
\cormark[1]



\credit{Conceptualization, Resources, Writing - Original Draft, Writing - Review \& Editing, Supervision, Project administration, Funding acquisition}

\author[address1,address2]{Yuan Xie}[style=chinese, orcid=0000-0003-3803-0929]
\ead{xieyuan@hccl.ioa.cn.cn}
\credit{Methodology, Software, Validation, Formal analysis, Investigation, Data Curation, Writing - Original Draft, Writing - Review \& Editing, Visualization}
\fnmark[1]
\fntext[fn1]{Co-first author.}

\author[address1,address2]{Wenchao Wang}[style=chinese]
\ead{wangwenchao@hccl.ioa.cn.cn}

\credit{Methodology, Investigation, Data Curation, Supervision}

\affiliation[address1]{organization={Key Laboratory of Speech Acoustics and Content Understanding, Institute of Acoustics, Chinese Academy of Sciences},
    addressline={No.21, Beisihuan West Road, Haidian District},
    postcode={100190},
    city={Beijing},
    country={China}}
    
\affiliation[address2]{organization={University of Chinese Academy of Sciences},
    addressline={No.80, Zhongguancun East Road, Haidian District}, 
    postcode={100190},
    city={Beijing},
    country={China}}


\cortext[cor1]{Corresponding author}

\fntext[fn1]{The source code of this paper could be obtained from https://github.com/xy980523/SmoothReg-LMR-DataPrune--underwater-acoustic-recognition.}



\begin{abstract}
Underwater acoustic target recognition is a challenging task owing to the intricate underwater environments and limited data availability. Insufficient data can hinder the ability of recognition systems to support complex modeling, thus impeding their advancement. To improve the generalization capacity of recognition models, techniques such as data augmentation have been employed to simulate underwater signals and diversify data distribution. However, the complexity of underwater environments can cause the simulated signals to deviate from real scenarios, resulting in biased models that are misguided by non-true data. In this study, we propose two strategies to enhance the generalization ability of models in the case of limited data while avoiding the risk of performance degradation. First, as an alternative to traditional data augmentation, we utilize smoothness-inducing regularization, which only incorporates simulated signals in the regularization term. Additionally, we propose a specialized spectrogram-based data augmentation strategy, namely local masking and replicating (LMR), to capture inter-class relationships. Our experiments and visualization analysis demonstrate the superiority of our proposed strategies.
\end{abstract}




\maketitle

\section{Introduction}
Underwater acoustic target recognition is a crucial component of marine acoustics~\citep{brooker2016measurement,xie2022underwater}, with widespread application in automating maritime traffic monitoring and noise source identification in ocean environmental monitoring systems~\citep{fillinger2010towards,xie2022adaptive}. The accurate recognition of targets based on radiated noise has high practical value. In recent years, research aimed at building robust underwater acoustic recognition systems has grown significantly due to the increasing demand~\citep{li2017denoising,ke2020integrated}. Classic paradigm converts signals into non-redundant acoustic features for subsequent recognition. For example, Das et al.~\citep{das2013marine} employed a cepstrum-based approach for marine vessel classification, Wang and Zeng~\citep{wang2014robust} used a bark-wavelet analysis combined with the Hilbert-Huang transform, and Chen et al.~\citep{chen2021underwater} used LOFAR (low frequency analysis recording) to reflect the power spectrum distribution and signals changes in time and frequency dimensions. However, Irfan et al.~\citep{irfan2021deepship} indicated that conventional methods based on low-dimensional acoustic features have limitations on large-scale data with a diversified feature space. They pointed out that high-dimensional representations of acoustic signals could better capture the inherent characteristics of target signals.

With the rapid development of deep learning~\citep{lecun2015deep} and the accumulation of underwater acoustic databases~\citep{irfan2021deepship,santos2016shipsear}, recognition algorithm based on deep learning continues to grow in popularity. Applying neural networks to learn from high-dimensional spectrograms, which contain rich time-frequency information, has become a prevalent paradigm. For instance, Zhang et al.~\citep{zhang2021integrated} used the short-time Fourier transform (STFT) amplitude spectrogram, STFT phase spectrogram, and bi-spectrogram features as inputs for the convolutional neural network (CNN)~\citep{krizhevsky2012imagenet}. In addition, spectrograms based on Mel filter banks~\citep{liu2021underwater,courmontagne2012time}, Gabor transform~\citep{ren2022ualf}, and wavelet transform~\citep{xie2022adaptive} are also widely applied in CNN-based underwater acoustic recognition. Compared to low-dimensional acoustic features, time-frequency-based spectrograms provide comprehensive information, while neural networks can adaptively capture the deep semantic information and inherent characteristics in spectrograms~\citep{liu2021underwater,xie2020time,zhu2021convolutional}.

Despite advances brought by neural network-based methods, the scarcity of data hinders comprehensive modeling of the complex underwater acoustic environment~\citep{irfan2021deepship,santos2016shipsear,gao2020recognition}. Limited data cause models to suffer from severe overfitting problems~\citep{xie2022underwater}, thus seriously hurting the performance and generalization ability of recognition models in real scenarios. To overcome the overfitting problem, researchers often augment training data by simulating underwater signals and generating noisy samples. Some researchers analyze the acoustic characteristics of underwater environments and manually design environment-related noisy samples (e.g., simulate different levels of marine turbulence~\citep{huang2019faster}), while some researchers diversify data distributions by generating virtual samples through generative adversarial networks (GANs)~\citep{gao2020recognition,yang2020gan}. However, according to our experiments (see Section 4), manual data augmentation and GAN-based augmentation could not consistently lead to improvements due to the inevitable deviations~\citep{gong2021eliminate} between simulations and real-world underwater environments. If data augmentation pushes the supplementary data further from the true distribution, the performance of recognition models may even degrade. To address these limitations, our work presents two effective training strategies for mitigating overfitting while avoiding the risk of getting influenced by low-quality noisy samples.

First, inspired by local shift sensitivity from robust statistics~\citep{huber2011robust}, we propose smoothness-inducing regularization as an effective alternative to traditional data augmentation methods. During training, simulated noisy samples are excluded from the direct loss calculation with ground truth, but are used to compute the regularization term based on the Kullback-Leibler (KL) divergence. Since noisy samples are not directly involved in the loss calculation, the model could minimize the risk of aggravating misjudgment even if noisy samples deviate from real-world scenarios. This property makes the decision boundary of the model smoother and reduces the model sensitivity to low-quality noisy samples, thus suppressing the risk of performance degradation.

The regularization strategy mitigates overfitting by constraining model training, which can be viewed as a conservative and moderate approach. Furthermore, inspired by the success of ``mixup'~\citep{zhang2017mixup} and ``cutmix''~\citep{yun2019cutmix} in pattern recognition tasks, we introduce a novel data augmentation strategy called local masking and replicating (LMR), which is designed specifically for time-frequency spectrogram-based recognition. LMR generates mixed spectrograms by randomly masking local patches in effective frequency bands and replicating patches in the same region from other spectrograms. Since the local regions in spectrograms contain line spectrum or modulation information within a certain time-frequency window, LMR can help the model enhance the ability to capture inter-class relationships from mixed spectrograms. This novel data augmentation strategy introduces a regularization effect that mitigates overfitting and improves the model's generalization to unseen data. 



In this work, we thoroughly evaluate the effectiveness of our proposed methods on three underwater ship-radiated noise databases: Shipsear~\citep{santos2016shipsear} (with a recognition accuracy of 83.45\%), DeepShip~\citep{irfan2021deepship} (with a recognition accuracy of 80.05\%), and a private dataset~\citep{ren2019feature} (with a recognition accuracy of 97.80\%). To demonstrate the stability and robustness of our methods, we present comprehensive visualizations for further analysis. The key contributions of this work are summarized as follows:

\begin{itemize}
\item This work experimentally reveals the limitations of traditional data augmentation techniques that rely on the quality of simulated samples.

\item We introduce a smoothness-inducing regularization strategy to alleviate overfitting and minimize the risk of performance degradation when noisy samples deviate from real-world scenarios.

\item We further propose a specialized spectrogram-based data augmentation strategy - LMR, that captures inter-class relationships and improves the model's generalization capacity.

\item We point out the incomparability of existing work and release our train/test split to establish a benchmark for follow-up work.

\end{itemize}

\section{Methodology}
This section presents an in-depth introduction to the concepts and implementation of our proposed recognition system, which is built upon two innovative techniques: smoothness-inducing regularization, local masking and replicating.

\subsection{Data Preprocessing and Feature Extraction}
In this work, we focus on the underwater acoustic signals detected by passive sonar arrays. During the preprocessing stage, array signal processing techniques, such as beamforming, are applied to enhance the signals in target directions while suppressing interference and noise from other directions.

\begin{figure*}
    \centering
    \subfigure[The overview of the preprocessing operations. The signal radiated by the target is collected by passive sonar.]{
        \begin{minipage}[b]{1\textwidth}
        \includegraphics[width=0.8\linewidth]{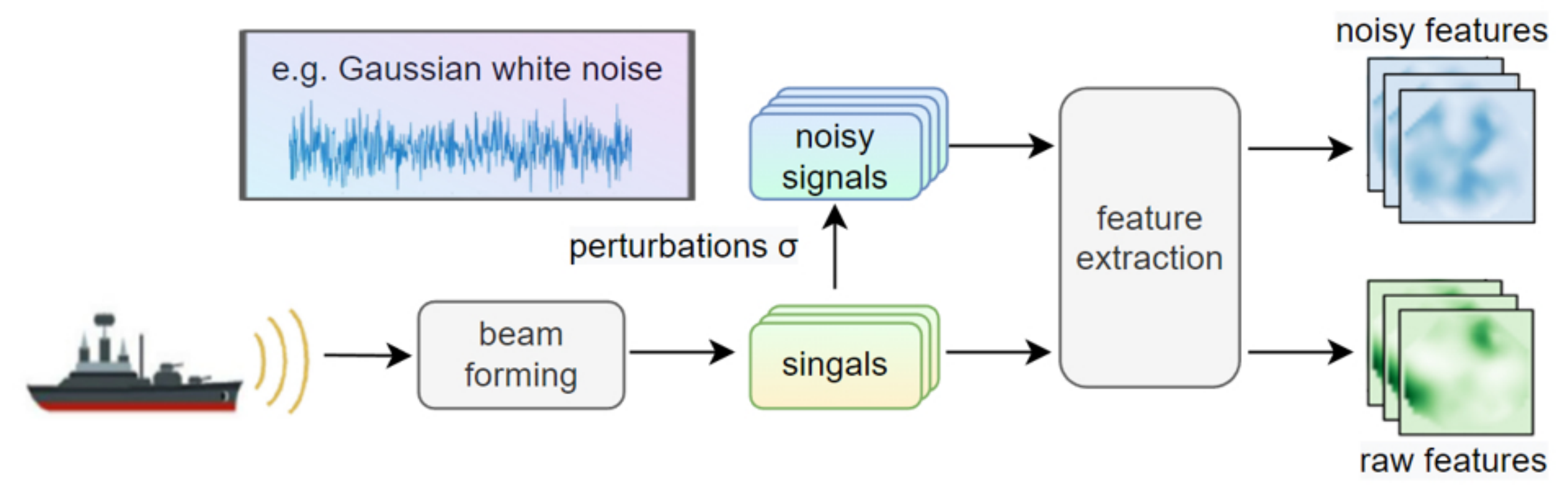}
        \centering
        \end{minipage}}

    \subfigure[Detailed steps of feature extraction. Grey boxes represent the operations and mauve boxes represent the extracted two-dimensional features.]{
        \begin{minipage}[b]{1\textwidth}
        \includegraphics[width=0.8\linewidth]{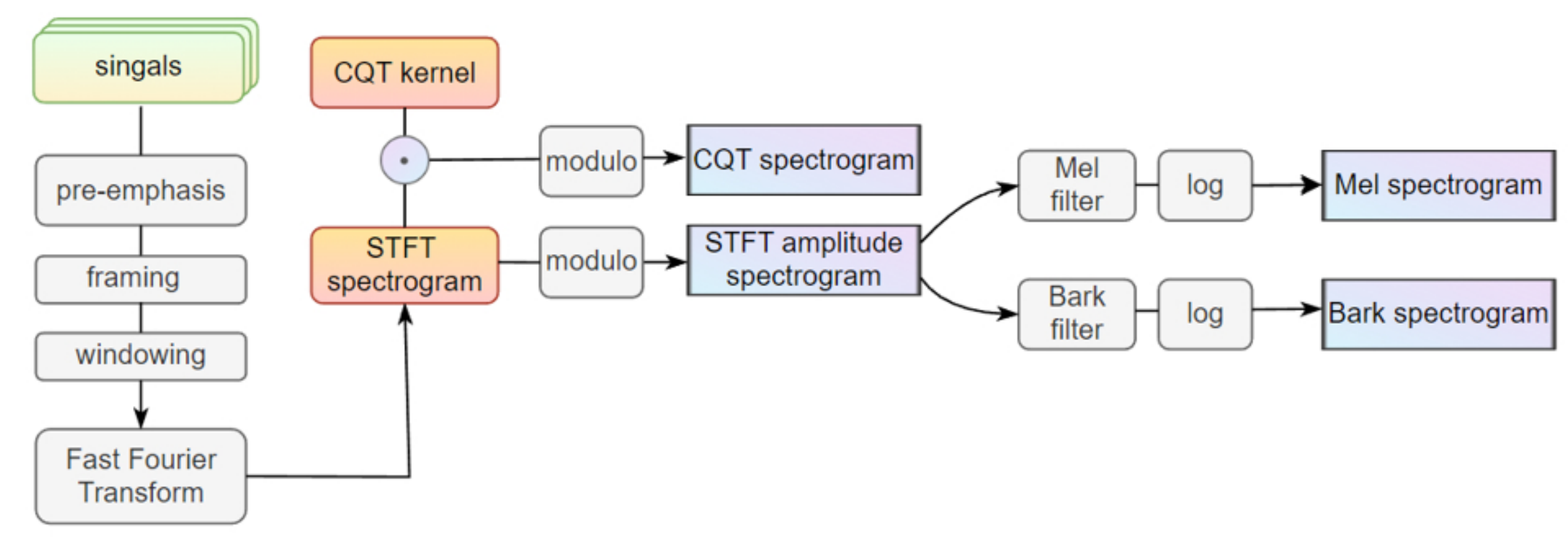}
        \centering
        \end{minipage}}
    \caption{The pipeline of our preprocessing process, along with the generation of noisy signals and the extraction of acoustic features.}
    \label{fig:pipeline}
    \vspace{-2px}
\end{figure*}

Figure~\ref{fig:pipeline}(a) provides an overview of the preprocessing operations. The process begins by implementing beamforming to enhance the signals in target directions, followed by simulating possible perturbations (e.g., background noise) or building generative adversarial networks to generate noisy signals. The subsequent feature extraction process transforms all signals into acoustic features. The features derived from the original signals are referred to as ``raw features'', while those obtained from the noisy signals are designated as "noisy features."

To validate the universality of our proposed strategies, we implement four feature extraction techniques in this work.  Figure~\ref{fig:pipeline}(b) presents a detailed process of our feature extraction process. We begin with computing the complex FFT spectrums through framing, windowing, and short-time Fourier transform. The real component is then extracted and integrated to obtain the STFT amplitude spectrogram. Subsequently, Mel (Bark) filter banks are applied to the framed FFT spectrums for filtering, as shown in Equation (1) where $f$ denotes frequency. Then, the filtered spectrums are converted to Mel (Bark) spectrograms using a logarithmic scale and the integrating operation.

\begin{equation}
\begin{aligned}
    Mel(f)
    &=2595\times log(1+\frac{f}{700}),\\
     Bark(f)
    &=6 \times arsinh(\frac{f}{600}).
\end{aligned}
\end{equation}

Furthermore, we obtain the CQT spectrogram through the constant Q transform. We convolve the framed FFT spectrums with the CQT kernel, which is a bank of bandpass filters logarithmically spaced in frequency. Equation (2) formalizes the $k$-th frequency component $f_k$, where the octave resolution is denoted by $b$, and the maximum and minimum frequencies to be processed are represented by $f_{max}$ and $f_{min}$, respectively.

\begin{equation}
    CQT(f_k)=2^{k/b} f_{min}. \quad f_{min}\leq f_k\leq f_{max}
\end{equation}

Then, the magnitude of the filtered spectrum is integrated to represent the CQT spectrogram.

\subsection{Backbone Model}
\begin{figure*}
    \centering
    \includegraphics[width=0.75\linewidth]{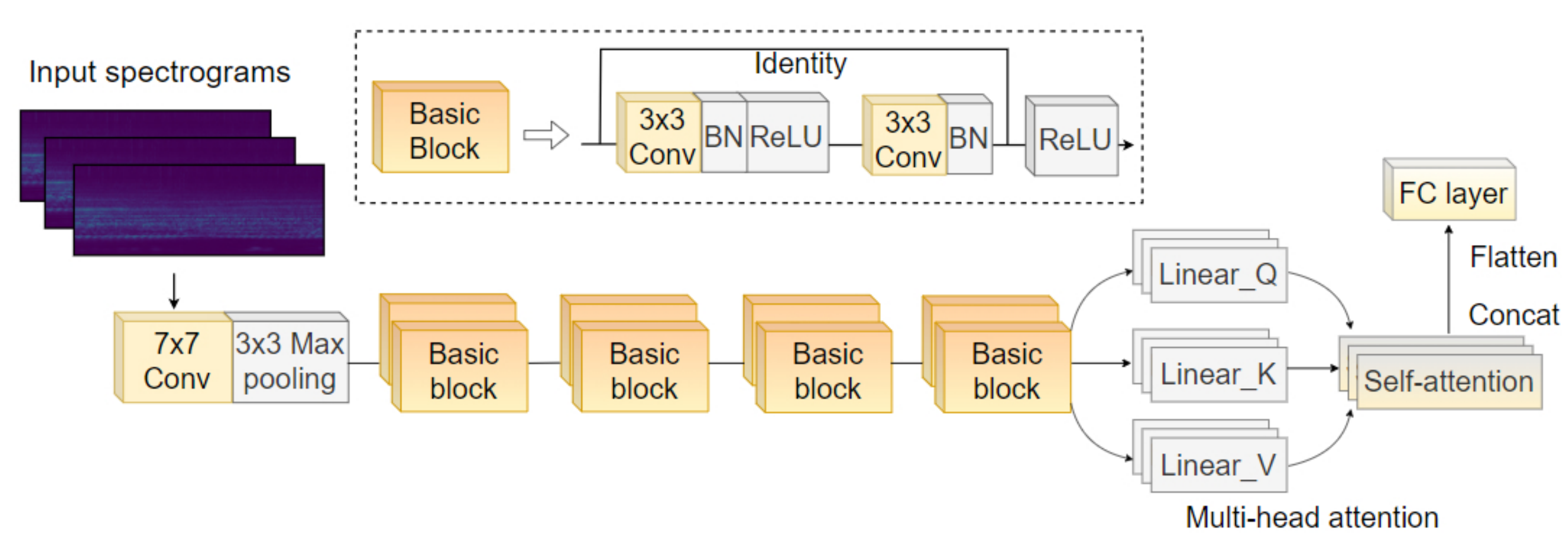}
    \caption{The structure of our backbone model: ResNet-18 with multi-head attention. ``Conv'' represents the convolutional layer, ``BN'' represents the two-dimensional batch normalization layer and ``FC'' represents the fully-connected layer.}
    \label{fig0}
    \vspace{-5px}
\end{figure*}

According to preliminary experiments (see Section 4.1), we adopt ResNet-18 with multi-head attention, an optimized convolutional neural network, as the backbone for our model. As shown in Figure~\ref{fig0}, our backbone model consists of a 7$\times$7 convolution layer stacked with a 3$\times$3 max-pooling layer with a stride of 2, followed by four residual layers and an attention pooling layer. Each residual layer contains the stack of two basic block layers, each including two 3$\times$3 convolution layers, two batch normalization layers, two ReLU layers, and a skip connection. The attention pooling layer implements a ``QKV'' (query-key-value) style multi-head attention~\citep{wang2018non}, which assigns different weights to regions of the feature map to focus on useful information, allowing the model to learn which features are important and should be emphasized in the recognition process. Then, the outputs of multi-head attention layers are concatenated and flattened into a one-dimensional representation. Finally, we use a task-related fully-connected layer to convert the representation to prediction results.

\subsection{Smoothness-inducing Regularization}
\begin{figure*}
    \centering
    \includegraphics[width=0.75\linewidth]{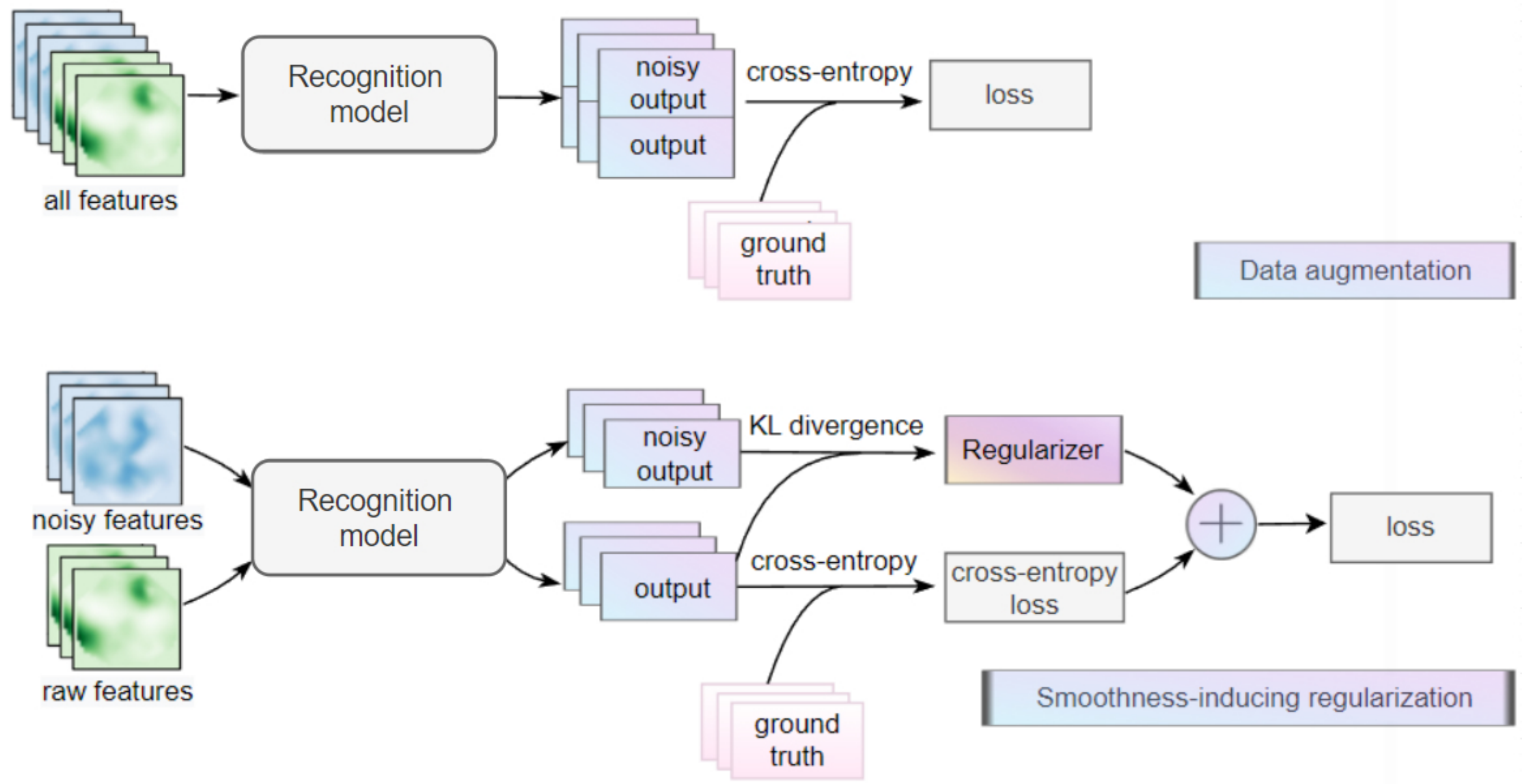}
    \caption{Comparison of the data augmentation and our smoothness-inducing regularization. For brevity, we omit the backpropagation of loss.}
    \label{fig3}
    \vspace{-5px}
\end{figure*}

We compare the traditional data augmentation method and our smoothness-inducing regularization during the training process in Figure~\ref{fig3}. Traditional data augmentation treats both raw and noisy features as training samples, while our smoothness-inducing regularization strategy prevents noisy samples from directly contributing to the loss calculation with the ground truth. This mitigates the reliance on the quality of noisy samples and leads to a more robust training process. Our methodology is as follows:

Denote raw features as $x_i$, noisy features as $\tilde{x_i}$ and corresponding labels as $y_i$ ($i = 1, 2 ...n$). The model $f(\cdot)$ is fed with paired samples ($x_i,y_i$) and ($\tilde{x_i},y_i$) simultaneously and outputs the logits $z_i=f(x_i)$ and $\tilde{z_i}=f(\tilde{x_i})$ after forward propagation. Inspired by previous work~\citep{jiang2019smart}, we apply the Kullback-Leibler (KL) divergence as the basis for the regularization term. Our optimization goal is to minimize the loss function:

\begin{equation}
\begin{aligned}
    \mathcal L 
    &=\mathcal L_{CE}(z_i,y_i) + \alpha \mathcal L_{Reg}(z_i,\tilde{z_i})\\
    &=\mathcal L_{CE}(z_i,y_i) + \alpha(\mathcal L_{KL}(z_i,\tilde{z_i})+ \mathcal L_{KL}(\tilde{z_i},z_i)),
\end{aligned}
\end{equation}

where $\mathcal L_{CE}$ represents the cross-entropy loss, and the weight coefficient $\alpha$ serves to adjust the weight of the regularization term, thus controlling the impact of smoothness-inducing regularization on the recognition model. We could set $\alpha=0$ to deactivate the smoothness-inducing regularization when it is not needed. The selection of the value of $\alpha$ can be found in Section 4.4.

$\mathcal L_{Reg}$ represents the regularization term based on KL divergence. KL divergence measures the ``distance'' between two discrete probability distributions. We feed the logits $z_i$ and $\tilde{z_i}$ into the softmax function and get the normalized probability distribution $p_i$ and $\tilde{p_i}$. The KL divergence-based term $\mathcal L_{KL}$ can be formulated as:

\begin{equation}
\begin{aligned}
    \mathcal L_{KL}(z_i,\tilde{z_i}) 
    &= \mathcal D_{KL}(p_i\Vert \tilde{p_i})
    &=  \frac{1}{n}\sum^{n}_{i=1} p_ilog\frac{p_i}{\tilde{p_i}},
\end{aligned}
\end{equation}

where $\mathcal D_{KL}(p_i\Vert \tilde{p_i})$ represents the KL divergence between $p_i$ and $\tilde{p_i}$. The regularization term in Equation (4) is minimized when the model does not significantly change predictions for perturbed samples ($p_i \approx \tilde{p_i}$). It reduces local sensitivity by encouraging similarity between the predictions for $x_i$ and its neighbors $\tilde{x_i}$, thus achieving a smoother decision boundary. This smoothness-inducing property can help improve the robustness of recognition models by reducing the sensitivity to noisy perturbations, contributing to consistent and reliable model performance.

\subsection{Local Masking and Replicating}
To further alleviate the overfitting problem and enhance the model's generalization ability, we propose a novel data augmentation strategy called local masking and replicating (LMR) that specializes in diversifying the distribution of training data for time-frequency spectrogram-based recognition. This technique is universally applicable and generates new training samples solely based on the raw data, eliminating the need for manual perturbations or adversarial samples.

\begin{figure*}
    \centering
    \includegraphics[width=0.75\linewidth]{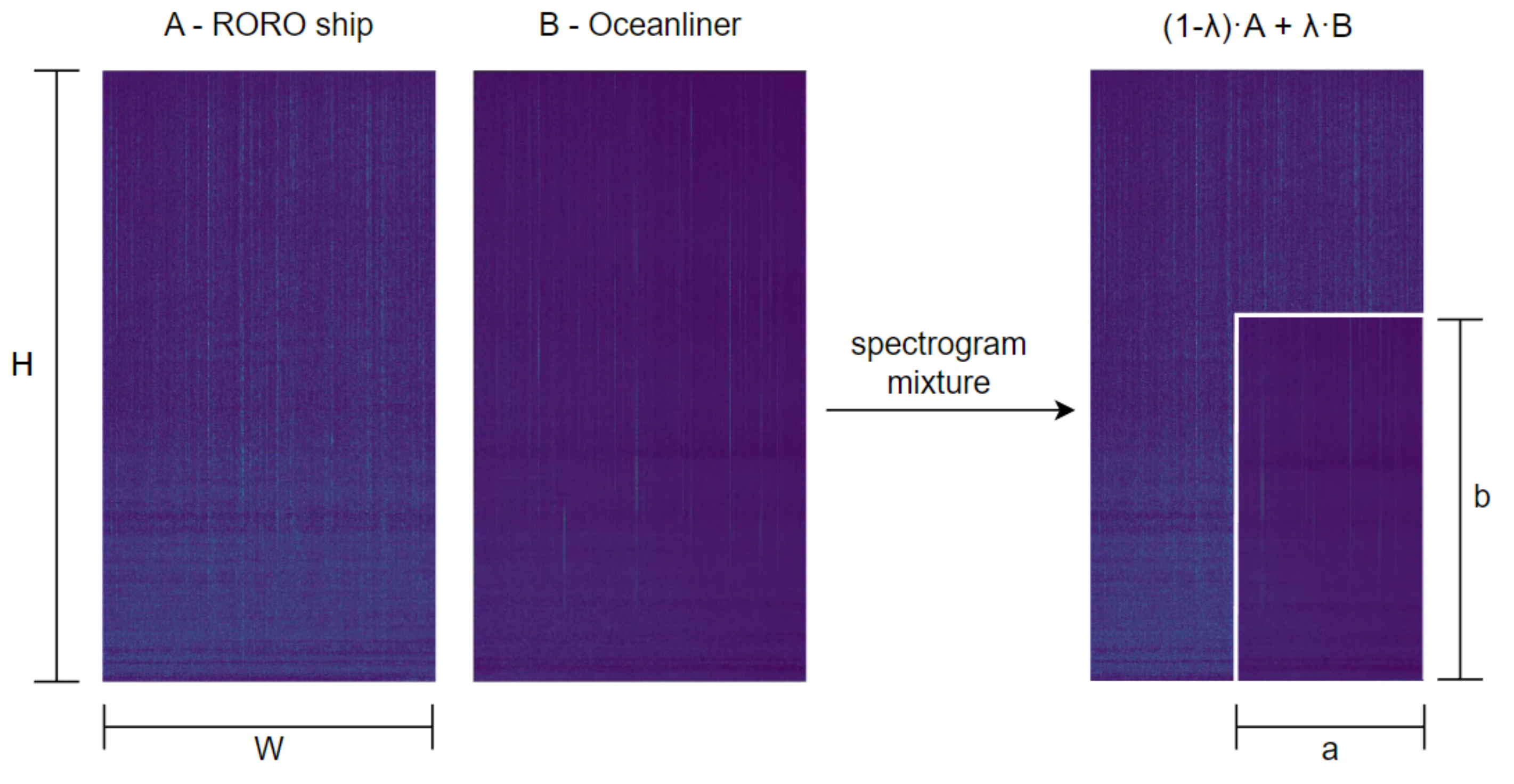}
    \caption{An example for local masking and replicating (LMR). The samples in the figure are selected from two 30-second segments in Shipsear.}
    \label{fig4}
    \vspace{-5px}
\end{figure*}

LMR is applied exclusively during the training phase. Before the training process, we analyze the effective frequency bands of signals and obtain input spectrograms with lower information redundancy by limiting the frequency bands. As depicted in Figure~\ref{fig4}, LMR randomly samples two input spectrograms of any category in the batch during training. One spectrogram randomly masks local patches, while patches of the same size and position from another spectrogram are replicated and inserted in the masked area of the former spectrogram. The mixed ``spectrogram'' serves as the input sample for training. We denote the width and height of input spectrograms as $W$ and $H$, respectively, and represent the removed region as an $a \times b$ patch where $a (a\in (0,W))$ and $b (b\in (0,H))$ represent the width and height of the masked and replicated regions. $a$ and $b$ are random values within their value range, which change dynamically with sampling. We denote the two sampled spectrograms and the mixed ``spectrogram'' as $x_i,x_j,\mathcal{X}_{ij}$, respectively. The model $f(\cdot)$ takes $\mathcal{X}{ij}$ as input and outputs $\mathcal{Z}{ij}=f(\mathcal{X}{ij})$ after feedforward. The coefficient $\lambda$ measures the proportion of mixture, where $\lambda=1-\frac{a\cdot b}{H\cdot W}$. The corresponding loss is formulated as:

\begin{equation}
\begin{aligned}
    \mathcal L_{LMR}
    &= \lambda \mathcal L_{ce}(\mathcal{Z}_{ij},y_i) + (1-\lambda) \mathcal L_{ce}(\mathcal{Z}_{ij},y_j),
\end{aligned}
\end{equation}

where $\mathcal L_{ce}$ denotes the cross-entropy loss. Notably, when LMR is activated, $\mathcal L_{LMR}$ will replace $\mathcal L_{CE}$ in Equation (3). Since the local regions in spectrograms contain line spectrum or modulation information within a certain time-frequency window, the mixed spectrograms $\mathcal{X}_{ij}$ may contain distinguishable features from different categories. The model can enhance the ability to capture inter-class relationships by directly learning from mixed spectrograms, thereby better distinguishing similar categories.

In addition, we activate LMR dynamically and intermittently during training. Each batch has a 50\% probability of being directly fed to the model (deactivate LMR), and a 50\% probability of implementing sampling and LMR operations. In this way, each sample has the opportunity to directly serve as input or participate in LMR. Compared with always activating LMR, this strategy can reduce computational consumption while diversifying the training inputs.

\subsection{Optimization objective}
When both of our proposed strategies are activated, our optimization objective can be represented as minimizing the loss function as follows:

\begin{equation}
\begin{aligned}
    \mathcal L
    &= \mathcal L_{LMR} + \alpha \mathcal L_{Reg} \\
    &= \mathcal L_{LMR} + \alpha [\mathcal L_{KL}(z_i,\tilde{z_i})+\mathcal L_{KL}(\tilde{z_i},z_i)] \\
    &= \lambda \mathcal L_{ce}(\mathcal{Z}_{ij},y_i) + (1-\lambda) \mathcal L_{ce}(\mathcal{Z}_{ij},y_j) + \\ &\alpha [\frac{1}{n}(\sum^{n}_{i=1} p_ilog\frac{p_i}{\tilde{p_i}} +\sum^{n}_{i=1} \tilde{p_i}log\frac{\tilde{p_i}}{p_i})].
\end{aligned}
\end{equation}

To minimize the optimization objective, the model should accurately capture the distinguishable characteristics of different classes while being insensitive to perturbations.

\section{Experiment Setup}
\subsection{Datasets}
In this work, we use three underwater ship-radiated noise datasets across different scales. The detailed information is presented in the subsequent paragraphs:

1. Shipsear~\citep{santos2016shipsear} is an open-source database of underwater recordings of ship and boat sounds. It currently comprises 90 records representing sounds from 11 vessel types, consisting of nearly 3 hours of recordings. To ensure sufficient data for the ``train, validation, test'' split, we select a subset of 9 categories (dredger, fish boat, motorboat, mussel boat, natural noise, ocean liner, passenger ship, ro-ro ship, and sailboat) for the recognition task.

2. DeepShip~\citep{irfan2021deepship} is an open-source underwater acoustic benchmark dataset, which consists of 47.07 hours of real-world underwater recordings of 265 different ships belonging to four classes (cargo, passenger ship, tanker, and tug).

3. Our private dataset~\citep{ren2019feature}- DTIL is a dataset collected from Thousand Island Lake, which contains multiple sources of interference. The dataset comprises 330 minutes of the speedboat and 285 minutes of the experimental vessel.

\subsection{Division of data}
In this work, we divide each signal into 30-second segments with a 15-second overlap. To prevent information leakage, we make sure that segments in the training set and the test set must not belong to the same audio track. This ensures that the reported accuracy reflects the recognition ability and generalization performance, rather than the memory capacity.

We note that many previous works on underwater acoustic recognition tasks do not provide their training/test split, which makes it challenging to make a fair comparison. Therefore, we release our carefully selected train/test split of Shipsear and DeepShip in Appendix B. During training, we randomly take 15\% of the data in the training set as the validation set. We hope that it could serve as a reliable benchmark for subsequent work pursuing fair comparisons. Table~\ref{tab0} displays the number of samples (30-second segments) in the training, validation, and test set for the three datasets.

\begin{table}[ht]
\normalsize
    \centering
    \caption{Number of samples (30-second segments) in the training, validation, test set for the three datasets.}
	\scalebox{1}{\begin{tabular}{llll}
		\hline
		  Split&Shipsear&DTIL&DeepShip\\
            \hline
            Training&400 &1655& 6742\\
            Validation&71&292 & 1190\\
            Test &  116  &354 & 2547\\
            \hline
            Total& 587   &2301& 10479\\ 
		\hline
        \label{tab0}
	\end{tabular}}
\end{table}

\subsection{Effective frequency bands}
The energy of the target signals primarily concentrates on specific frequency ranges. To facilitate the replicated area in LMR containing valid information, we take the effective frequency band as a replacement for the full-frequency band during analyzing underwater signals. Moreover, time-frequency transformation within the bandwidth could reduce the redundancy of acoustic features, thus enhancing recognition performance while reducing time and hardware consumption. The distribution of energy in target signals varies across different frequency bands for the three datasets, so we set different effective frequency bands for each (see Table~\ref{tab1}). Details regarding the effective frequency band selection could be found in Appendix A. Notably, according to the Nyquist theory, the upper limit of the effective frequency band must be less than half of the sample rate.

\subsection{Parameters Setup}
Regarding framing, we conduct experiments to investigate the effect of frame length on recognition results, which is shown in Appendix C. In this work, we set the frame length as 50ms and the frame shift as 25ms. Furthermore, we set the number of Mel or Bark filter banks to 300 as default. For the CQT kernel, we set the octave resolution to 30 and set the maximum(minimum) frequency according to the boundary of effective frequency band for each dataset. When extracting the CQT spectrogram, we set the time resolution to 30 as default. The dimensions of the various acoustic features are shown in Table~\ref{tab1}.

\begin{table*}[ht]
\normalsize
    \centering
    \caption{Information of the three datasets and corresponding feature dimensions. ``SR'' represents the sampling rate and ``Dim'' represents the dimension of features.}
	\scalebox{0.85}{\begin{tabular}{lccllcc}
		\hline
		  Dataset  & Duration (hours) & SR (Hz) & Efficient band (Hz)&
            STFT Dim&Mel/Bark Dim&CQT Dim\\
            \hline
            Shipsear  & 2.94 & 52734 & 100-26367&
            1199,1318&1199,300&899,340\\
            DTIL  & 10.25 & 17067 & 100-2000&
            1199,100&1199,300&899,229\\
            Deepship & 47.07 & 32000 & 100-8000&
            1199,400& 1199,300&899,289\\
		\hline
        \label{tab1}
	\end{tabular}}
\end{table*}

During training, we employ the AdamW~\citep{loshchilov2017decoupled} optimizer. We set the learning rate to 5e-4, batch size to 64, and weight decay to 1e-5 for all experiments. All models are trained for 100 epochs on a single A40 GPU.

\section{Results and analysis}
For the multi-class recognition problem addressed in this work, we uniformly adopt the accuracy rate as the evaluation metric. The accuracy rate is calculated by dividing the number of correct predictions by the total number of predictions. It provides an intuitive measure of the model's ability to distinguish between different classes accurately. Besides, since the number of complete audio files in the test set is limited, several groups of experiments yield the same file-level accuracy. Therefore, we report results at the segment level (30 seconds) rather than the file level.

In this section, we validate the effectiveness of our proposed methods on three datasets and conduct detailed comparison experiments to separately verify the gains brought by smoothness-inducing regularization and LMR. To better demonstrate the superiority of our proposed methods, we visualize our results by confusion matrix heat map and gradient-weighted class activation mapping (CAM)~\citep{selvaraju2017grad}. An introduction to the concept of the confusion matrix heat map and CAM can be seen in Section 4.2.

\subsection{Main results}
\begin{table*}[ht]
\normalsize
    \caption{\label{tab2} Main results across various datasets and features. ``smooth reg'' represents smoothness-inducing regularization for short. The underlined part indicates a decrease in performance compared to the baseline.}
        \scalebox{0.85}{
	\begin{tabular}{lllccc}

        \hline
	Model&Features & Methods & Shipsear & DTIL&DeepShip\\
        \hline
        Random forest (RF)& STFT spec& - &   
        59.33 &   82.56  &  66.71  \\
        Support vector machines (SVM) & STFT spec & -  &   
        65.65&   88.74  &   69.24\\
        FCN(1-d convolution)& STFT spec& - &   
        72.56&   91.67  &  72.98  \\
        MobileNet-v3 small& STFT spec& -  &   
        70.02&   90.54  &   70.45\\
        ResNet-18& STFT spec& - &   
        75.16 &   94.98  &   74.02 \\
        SE-ResNet-18& STFT spec& -  &   
        71.59&   92.25  &   71.36\\
        ResNet-18+linear attention& STFT spec& - &   
        74.99&    95.61 &  74.59  \\
        ResNet-18+multi-head attention& STFT spec& - & \textbf{75.24} & \textbf{95.93}& \textbf{74.68} \\
 
	\hline
	ResNet-18+multi-head attention& STFT spec & - & 75.24 & 95.93& 74.68 \\
        && +noise aug& 78.45 & 96.05 & 75.07  \\
        && +GANs aug& 76.05 & 96.14 & 75.98  \\
        && +smooth reg (noise)&  81.90 & 97.74& 76.38\\
        && +LMR& 82.76 & 96.61 & 78.40 \\
	&& +smooth reg, LMR&  \textbf{82.97} & \textbf{97.80} & \textbf{78.56}  \\

        \hline
        ResNet-18+multi-head attention&Mel spec & - & 77.14 &95.48& 74.85 \\
        && +noise aug& 78.72 & 95.58 & 75.27  \\
        && +GANs aug& 79.68 & \underline{95.45} & 75.79  \\
        && +smooth reg (noise)&82.76 & 95.76& 77.05\\
        && +LMR& 82.76 & 96.33 & 75.38 \\
	&& +smooth reg, LMR& \textbf{83.45}  & \textbf{96.50}  & \textbf{77.12}  \\

        \hline
        ResNet-18+multi-head attention&Bark spec & - &72.86 & 96.30& 75.15 \\
        && +noise aug& 75.86 & 96.33 &  \underline{74.18} \\
        && +GANs aug& 74.59 & 96.65 & \underline{75.08}  \\
        && +smooth reg (noise)& 77.72& 96.76&75.75\\
        && +LMR&\textbf{79.31} & 96.61 & 77.46 \\
	&& +smooth reg, LMR& 79.25 & \textbf{96.93} & \textbf{77.80}  \\

        \hline
        ResNet-18+multi-head attention&CQT spec & - & 73.33 & 96.48& 77.82 \\
        && +noise aug& 74.41 & 97.18 &  77.86 \\
        && +GANs aug& 74.43 & 96.89 &  \underline{77.12} \\
        && +smooth reg (noise)&75.86&97.18 &78.25\\
        && +LMR& 81.03 & 97.18 & 78.97\\
	&& +smooth reg, LMR& \textbf{82.76}  & \textbf{97.56} & \textbf{80.05} \\
        \hline
    \end{tabular}}
\end{table*}

First, we conduct preliminary experiments on the model backbone (see the top part of Table~\ref{tab2}). To ensure a fair comparison, we uniformly use the STFT spectrogram without any augmentation and regularization. We implement several baselines~\citep{howard2019searching,he2016deep,hu2018squeeze} that perform well in underwater acoustics and observe that the ResNet-based model has significant performance advantages over traditional methods (RF, SVM) and other convolutional neural networks (FCN, MobileNet). Among them, ResNet with multi-head attention achieves the best recognition accuracy on the three datasets. The application of multi-head attention mechanism allows the model to flexibly focus on effective information and avoid being influenced by redundant parts. Given its superior and stable performance, we choose ResNet with multi-head attention as the unified model backbone for all experiments in this work. A detailed description of model structures is available in Section 2.2 and Appendix D.


We then focus on the comparison experiments between various data augmentation and regularization strategies (see the bottom part of Table~\ref{tab2}). We implement two data augmentation strategies as the baseline methods: adding Gaussian white noise with the signal-to-noise ratio of 5-30dB (noise aug), and generating adversarial samples with vanilla GAN (GANs aug). To enable a fair comparison, when implementing smoothness-inducing regularization, we copy the parameter settings of ``noise aug'' to generate the same noisy samples. To avoid the influence of the amount of training data on the results, we ensure that all three augmentation strategies generate noisy samples equal in number to raw training samples. That is, the number of total training samples is doubled after augmentation.

As depicted in Table~\ref{tab2}, we find that classic data augmentation methods cannot consistently bring boosts and sometimes even have a negative effect (see the underlined results in Table~\ref{tab2}). Especially for the model based on Bark spectrogram on Deepship, the employment of ``noise aug'' reduces the recognition accuracy by 0.97\%. It shows that recognition models face the risk of performance degradation when noisy samples deviate from real-world scenarios. Models may learn biases from low-quality samples if they treat manually added perturbations as the inherent property of ship-radiated noise. Compared to classic data augmentation methods, our proposed regularization strategy consistently brings promising and stable improvement with minimal risk of performance degradation. As a regularization method, it avoids noisy samples from directly participating in model training, resulting in less negative impact.

Moreover, LMR achieves surprising results on all datasets. The performance gains from LMR sometimes even outweigh the gains obtained by smoothness-inducing regularization. Especially on Shipsear, LMR can bring 5.62\% $\sim$ 7.70\% accuracy improvement to models based on different features. It highlights the efficacy of constructing mixed samples to capture inter-class relationships during training. We also observe that the combination of smoothness-inducing regularization and LMR can further improve performance, indicating good compatibility between the two strategies. 

\subsection{Visualization}
\begin{figure*}
    \centering
    \subfigure[Confusion matrix heat maps for Shipsear.  There are nine categories: dredger, fish boat, motorboat, mussel boat, natural noise, ocean liner, passenger ship, ro-ro ship and sailboat.]{
        \begin{minipage}[b]{1\textwidth}
        \includegraphics[width=0.75\linewidth]{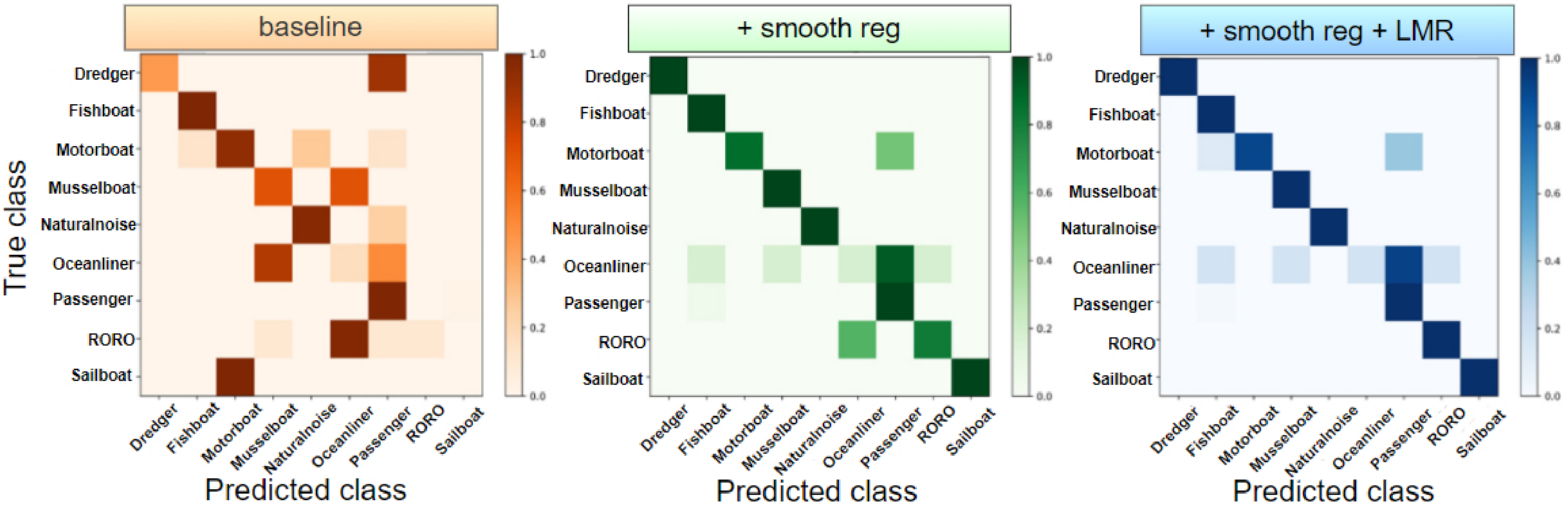}
        \centering
        \end{minipage}}

    \subfigure[Confusion matrix heat maps for DeepShip. There are four categories: cargo, passenger ship, tanker, tug.]{
        \begin{minipage}[b]{1\textwidth}
        \includegraphics[width=0.75\linewidth]{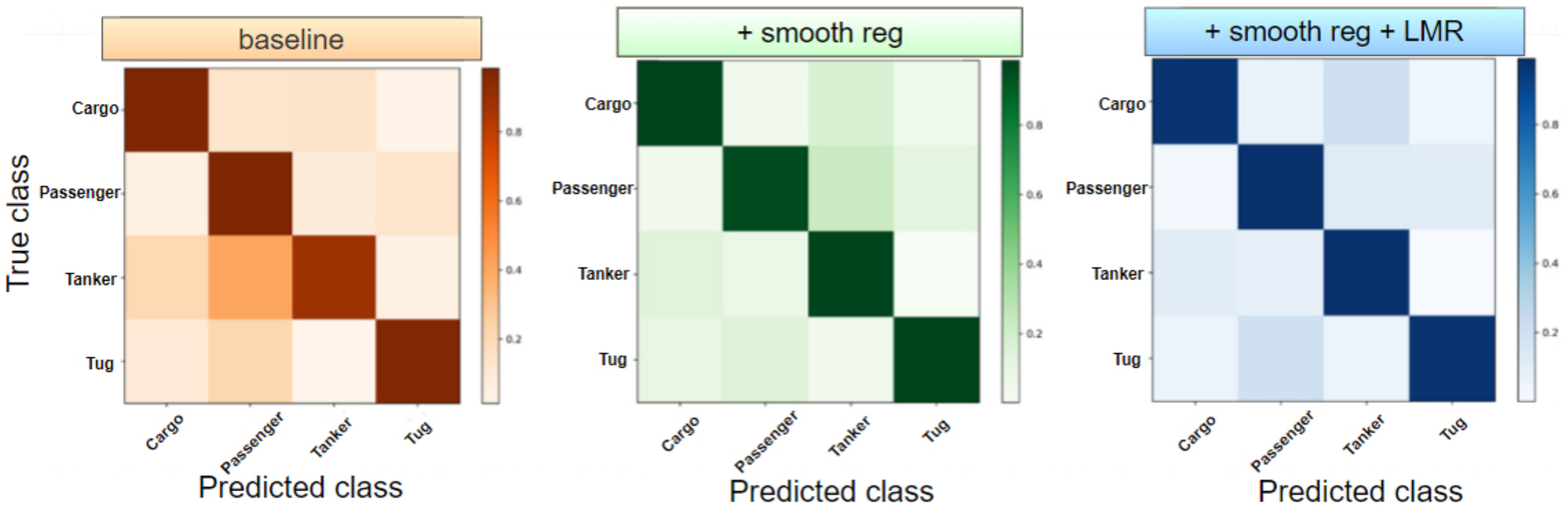}
        \centering
        \end{minipage}}
    \caption{Confusion matrix heat map for two datasets. The three subplots represent: 1. baseline; 2. with smooth reg; 3. with smooth reg and LMR. ``smooth reg'' represents smoothness-inducing regularization for short.}
    \label{fig5}
    \vspace{-5px}
\end{figure*}

This subsection aims to provide a visual representation of the results to facilitate comprehension and interpretation. We present the confusion matrix heat map for the recognition results and use the gradients in back-propagation as weights to draw the class activation mapping.

The confusion matrix compares the true labels (vertical axis) to the predicted labels (horizontal axis), and the confusion matrix heat map (see Figure~\ref{fig5}) is the graphical representation of the confusion matrix, where the values in the matrix are represented as colors. The color scale used in a heatmap ranges from light (representing low values) to dark (representing high values). Figure~\ref{fig5} shows that our proposed strategies improve the recognition accuracy of all categories. Take the results on Shipsear as an example, the baseline method often misclassifies dredgers, ocean liners, ro-ro ships, and completely fails to identify sailboats. After applying our smoothness-inducing regularization and LMR augmentation, the recognition accuracy for dredgers, ro-ro ships, and sailboats reaches 100\%, which is a very satisfactory improvement.

\begin{figure*}
    \centering
    \includegraphics[width=0.75\linewidth]{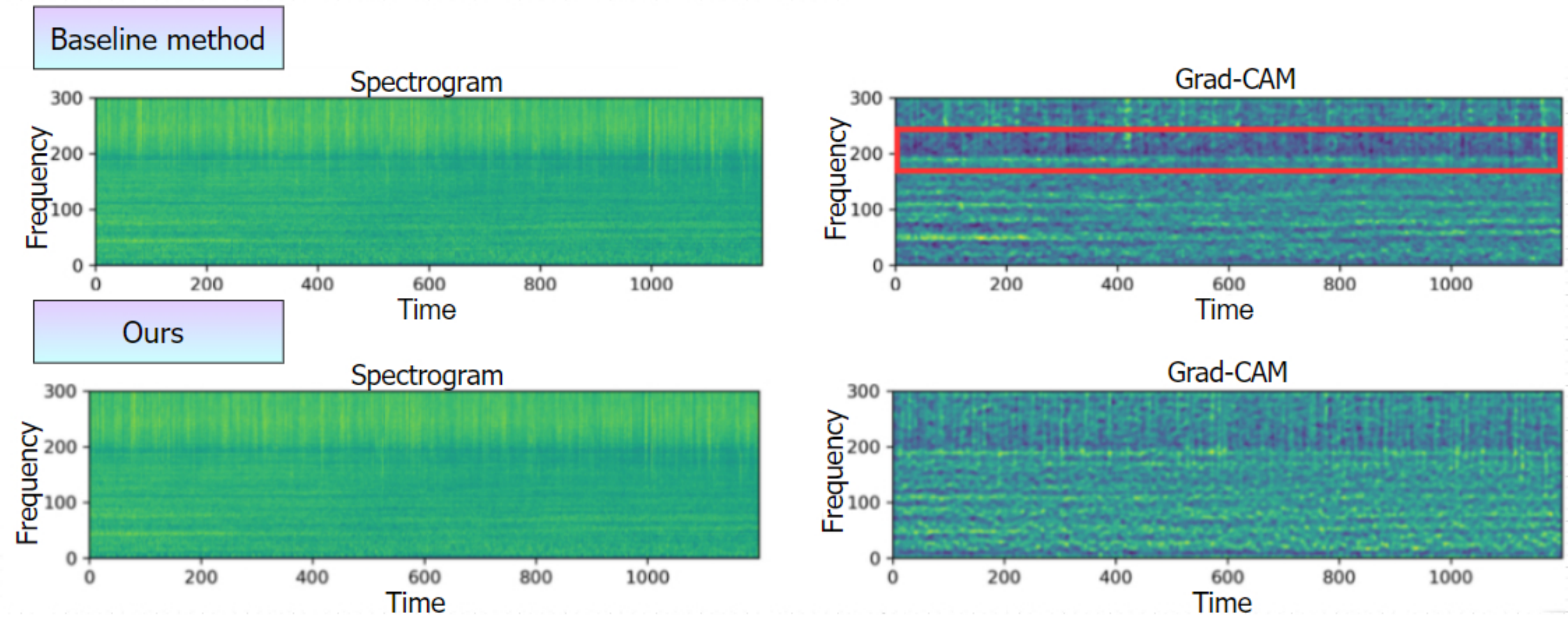}
    \caption{The Mel spectrogram and the corresponding CAM heat map for the baseline (top) and our method (bottom). We take the 30-second segment of a ro-ro ship in Shipsear as the example.}
    \label{fig7}
    \vspace{-5px}
\end{figure*}

Then, we generate the class activation mapping (CAM) by using the gradients in back-propagation as weights. The CAM is a graphical representation of the input spectrograms, with color indicating the impact of a region on the network's decision. Dark colors indicate a crucial role in the decision-making process. The CAM heat map faithfully highlights important regions in spectrograms for deep model prediction in recognition tasks. By analyzing the CAM in Figure~\ref{fig7}, we observe that our proposed optimizations enable models to emphasize global information, thus avoiding bias towards specific local information (e.g., see the CAM for the baseline model in Figure~\ref{fig7}, where the vertical scale is 200.). The avoidance of local bias can enhance recognition performance.

In addition, we display the four spectrogram-based features along with their corresponding CAM heat map in Appendix E. Our purpose is to show the critical components of different features that influence network decisions.

\subsection{Sensitivity to perturbations}
\begin{table*}[ht]
\normalsize
    \caption{\label{tab3} Performance of Mel-spectrogram-based models on the test set with different SNRs. We do not employ LMR to control interference factors.}
    \centering
	\scalebox{0.85}{
	\begin{tabular}{lcccc}
        \hline
	Method & Range of SNR & Shipsear & DTIL&DeepShip\\
	\hline
        Noise augmentation & - & 78.72 & 95.58& 75.27 \\
          & 5 $\sim$ 30 & 78.52 & 95.50& 75.19 \\
          & -5 $\sim$ 20 & 76.05 & 94.46& 73.36 \\
          & -15 $\sim$ 10 & 71.69 & 92.45& 68.67 \\
        \hline
        smoothness-inducing regularization (noise)& - & 82.76 & 95.76& 77.05 \\
          & 5 $\sim$ 30 & 82.66 & 95.63& 77.01 \\
          & -5 $\sim$ 20 & 81.95 & 94.78& 76.66 \\
          & -15 $\sim$ 10 & 79.69 & 93.70& 73.03 \\
        \hline
	\end{tabular}}
\end{table*}

In this set of experiments, we add white Gaussian noise with different signal-to-noise ratios (SNRs) to the test data. A lower SNR indicates a larger magnitude of the noise perturbation. We use trained models (as mentioned above, trained with 5$\sim$30dB noise perturbations) to evaluate the test data with different SNRs. We aim at exploring whether models based on smoothness-inducing regularization are less sensitive to perturbations. As illustrated in Table~\ref{tab3}, the model trained with manual noise augmentation maintains a performance drop of less than 3\% when the SNR of the test data is not lower than -5dB, whereas the recognition accuracy sharply decreases by 7.03\% when tested on data with -15$\sim$10dB SNR. By contrast, the model based on smoothness-inducing regularization exhibits consistent performance even under substantial perturbations. When the SNR of the test data reaches -15dB, the performance of our model only drops by 3.07\%, and the recognition accuracy remains at approximately 80\%. Our model can benefit from the smoother decision boundary brought about by the regularization strategy. Therefore, we conclude that models employing smoothness-inducing regularization are less sensitive to perturbations and can exhibit superior robustness against perturbations.

\subsection{Selcetion of regularization coefficient}
\begin{table}[ht]
\normalsize
    \caption{\label{tab4} Experiments with different regularization coefficient $\alpha$ across various acoustic features. We do not employ LMR to control interference factors.}
    \centering
	\scalebox{0.9}{
	\begin{tabular}{lcccc}
        \hline
	Features & $\alpha$ & Shipsear & DTIL&DeepShip\\
	\hline
	STFT spec & 0 & 75.24 & 95.93& 74.68 \\
        & 0.5& \textbf{81.90} & 97.18& 75.83 \\
        & 1& 81.47 & 97.05& 75.52 \\
	& 2& 78.45 & \textbf{97.74}& \textbf{76.38} \\
        
        \hline
        Mel spec & 0 & 77.14 & 95.48& 74.85 \\
        & 0.5& 81.14 & \textbf{95.76}&\textbf{77.05} \\
        & 1& 81.41 & 95.48& 75.79\\
	& 2& \textbf{82.76} & 95.20& 75.13 \\
        \hline
        
        Bark spec & 0 & 72.86 & 96.30& 75.15 \\
        & 0.5& \textbf{77.72} & 96.48& 75.40 \\
        & 1& 76.28 & \textbf{96.76}& \textbf{75.75} \\
	& 2& 76.00 & 96.20& 74.93\\
        \hline
        CQT spec & 0 & 73.33 & 96.48& 77.82 \\
        & 0.5& 75.28 & \textbf{97.18}& 77.86 \\
        & 1& \textbf{75.86} & 96.89& 78.17 \\
	& 2&75.41 & 96.61& \textbf{78.25} \\
        \hline
	\end{tabular}}
\end{table}

In this subsection, our focus is on selecting the appropriate regularization coefficient $\alpha$. As shown in Table~\ref{tab4}, the value of $\alpha$ has a noticeable impact on the results. Taking the model based on STFT features on shipsear as an example, setting $\alpha$ as 0.5 achieves a recognition accuracy rate of 81.90\%. By contrast, setting $\alpha$ as 2 can only get an accuracy rate of 78.45\%, with a gap of 3.45\%. Therefore, it is important to set a suitable $\alpha$ to make the smoothness-inducing regularization reach the best performance for different databases and acoustic features. If $\alpha$ is set too low, the regularization constraint is insufficient, and the model may still suffer from overfitting. Conversely, if $\alpha$ is set too high, the regularization may become overly aggressive and impede model learning. Typically, selecting $\alpha$ from the range of 0.5, 1, or 2 yields favorable outcomes. When confronted with diverse data and features, it is necessary to experiment with different values to obtain the optimal $\alpha$.

\subsection{Capture inter-class relationships: LMR vs mixup}
\begin{table}[ht]
\normalsize
    \caption{\label{tab5} Comparison experiments on LMR and mixup. The underlined part indicates a decrease in performance compared to the baseline. We do not employ regularization to control interference factors.}
    \centering
	\scalebox{0.8}{
	\begin{tabular}{lcccc}
        \hline
	Features & Augmentation & Shipsear & DTIL&DeepShip\\
	\hline
	STFT spec & - & 75.24 & 95.93& 74.68 \\
        & mixup & 79.31 & \underline{95.76} &74.83\\
        & LMR & \textbf{82.76} & \textbf{96.61} &\textbf{78.40} \\
        \hline
        Mel spec & - & 77.14 & 95.48& 74.85 \\
        & mixup & \underline{76.72} & 96.05 & \textbf{75.42} \\
        & LMR & \textbf{82.76} & \textbf{96.33} & 75.38 \\
        \hline
        Bark speck & - & 72.86 & 96.30&75.15 \\
        & mixup & 75.86 & 96.51 & \underline{73.77} \\
        & LMR & \textbf{79.31} & \textbf{96.61} & \textbf{77.46} \\
        \hline
        CQT spec & - & 73.33 & 96.48& 77.82 \\
        & mixup & 79.31 & \textbf{97.18} & 78.48 \\
        & LMR & \textbf{81.03} & \textbf{97.18} & \textbf{78.97} \\
        
        \hline
	\end{tabular}}
\end{table}

For pattern recognition tasks, mixup~\citep{zhang2017mixup} is a popular data augmentation technique that can capture inter-class relationships. It performs a linear mixture operation between the input and its corresponding label to generate virtual input samples. In this set of experiments, we implement comparison experiments between LMR and mixup. As indicated by the results in Table~\ref{tab5}, LMR provides greater gains for models based on four different acoustic features. Notably, on the data-scarce Shipsear dataset, LMR improves accuracy by 2.86\%$\sim$9.04\%. Additionally, we observe that LMR is more stable in most cases. Mixup may cause a slight accuracy drop (see the underlined results in Table~\ref{tab5}), whereas LMR does not show any risk of performance degradation in our experiments.

According to our analysis, since the stripes in the spectrogram contain line spectrum and periodic modulation information, mixing them linearly may result in aliasing or distortion, thus hurting the performance. The aliasing of the line spectrum or periodic stripes may cause a shift in the signal's frequency or period. Therefore, we argue that ``mixup'' is not ideal for spectrogram-based acoustic recognition. Conversely, our proposed LMR is specially designed for spectrogram-based recognition. Regions from different spectrograms are strictly kept separate to prevent aliasing. LMR enables virtual samples to capture inter-category relationships without the risk of aliasing, making it a more suitable approach for spectrogram-based acoustic recognition.

\section{Conclusion}
Through our experimental investigations, we have uncovered the limitations of traditional data augmentation techniques that heavily rely on signal quality and augmentation methods. Based on the observation, we propose innovative regularization and augmentation techniques that are less reliant on the quality of noisy samples. Our proposed smoothness-inducing regularization allows models to learn robust knowledge from noisy samples while minimizing biases toward low-quality supplementary training samples. Additionally, we introduce a specialized spectrogram-based augmentation strategy, called local masking and replicating, which enhances the diversity of the training data distribution and captures inter-class relationships between samples. Through comprehensive experiments and visual analyses, we demonstrate the effectiveness of our proposed techniques.

However, our experiments also reveal certain limitations of our proposed methods. First, while smoothness-inducing regularization performs well on the data-scarce Shipsear dataset, it only provides limited improvements when applied to the data-abundant DeepShip dataset. This indicates that the regularization technique can only mitigate the overfitting problem, rather than enhance the model's performance upper bound. Furthermore, the application scenarios of LMR are relatively limited. Despite its efficacy in spectrogram-based multi-class recognition, LMR can sometimes have negative effects on other tasks, such as detection tasks.

\section*{Acknowledgements}
This research is supported by the IOA Frontier Exploration Project (No. ZYTS202001) and Youth Innovation Promotion Association CAS.


\section*{Appendix}
\subsection*{A. Selection of effective frequency bands}
\begin{table}[htbp]
\normalsize
    \caption{\label{tabA} Selection of effective frequency bands. The upper limit of their effective frequency bands varies due to the different sampling rates.}
    \centering
	\scalebox{1}{
	\begin{tabular}{lccc}
        \hline
	frequency bands(Hz) & Shipsear & DTIL&DeepShip\\
	\hline
	100-2k & - & \textbf{95.93} & - \\
        100-4k&  - & 95.08& 74.37 \\
        100-8k& 60.69 & 94.52& \textbf{74.68} \\
	100-12k& 64.14 & -& 74.53 \\
        100-16k& 70.90 & -& 73.66 \\
        100-24k& 74.34 & -& - \\
        100-26.367k& \textbf{75.24} & -& - \\
        \hline
	\end{tabular}}
\end{table}

Table~\ref{tabA} presents the recognition results obtained using different frequency bands. The input feature for all experiments is the STFT spectrogram, and the model backbone for all experiments is ResNet with multi-head attention. No regularization or augmentation technique is applied. The frequency band with the best performance is selected as the effective frequency band for the three databases, as shown in Table~\ref{tab1}.

\subsection*{B. Train/Test split for Shipsear and DeepShip}
We find that almost all previous works on underwater acoustic recognition tasks do not release their split of train/test sets, which makes it challenging to make a fair comparison. In this work, we release our carefully selected train/test split of Shipsear and DeepShip (see Table~\ref{tabB1},~\ref{tabB2}). We believe that it could serve as a reliable benchmark for subsequent work pursuing fair comparisons\footnote{The train-test split for Shipsear is also released at https://github.com/xy980523/ShipsEar-An-Unofficial-Train-Test-Split}.

\begin{table}[ht]
\normalsize
    \caption{\label{tabB1} Train/Test split for Shipsear. The ``ID'' in the table refers to the ID of the .wav file in the dataset.}
    \centering
	\scalebox{0.67}{
	\begin{tabular}{lll}
        \hline
	Category& ID in Training set & ID in Test set\\
	\hline
	dredger&80,93,94,96& 95 \\
    fish boat&73,74,76&75 \\
    motorboat&21,26,33,39,45,51,52,70,77,79&27,50,72 \\
    mussel boat&46,47,49,66&48 \\
    natural noise&81,82,84,85,86,88,90,91&83,87,92\\
    ocean liner&16,22,23,25,69&24,71\\
    passenger ship&06,07,08,10,11,12,14,17,32,34,36,38,40,&9,13,35,42,55,62,65\\
    &41,43,54,59,60,61,63,64,67& \\
    ro-ro ship&18,19,58&20,78\\
    sailboat&37,56,68&57\\
        \hline
	\end{tabular}}
\end{table}

\begin{table}[ht]
\normalsize
    \caption{\label{tabB2} Train/Test split for DeepShip. The ``ID'' in the table refers to the ID of the .wav file in the dataset.}
    \centering
	\scalebox{0.67}{
	\begin{tabular}{lll}
        \hline
	Category& ID in Training set & ID in Test set\\
	\hline
    cargo& Else & 01,02,04,05,18,30,32,35,40,48,56,62,\\
    & &63,67,68,72,74,79,83,91,92,93,95,97,\\
    & &100,104\\

    \hline
    passenger ship& Else&02,04,05,07,11,15,17,19,20,23,30,31,\\
    &&37,45,46,52,53,60,61,62,64,67,68,70,\\
    &&75,76,77,84,86,91,101,106,113,117,122,\\
    &&125,129,130,134,135,142,144,152,157,\\
    &&159,161,167,168,177,179,187,188,189\\

    \hline
    tanker& Else&02,03,04,07,08,13,14,15,19,22,25,28,\\
    &&35,37,46,58,62,71,73,79,82,84,88,89,\\
    &&92,99,106,115,118,124,126,127,131,134,\\
    &&141,144,147,151,153,156,158,167,171,\\
    &&178,179,185,186,190,192,193,201,205,\\
    &&213,217,228,233\\
    \hline
    tug& Else& 07,08,18,20,24,25,27,29,32,33,37,39,\\
    &&40,44,45,56,59,70\\
        \hline
	\end{tabular}}
\end{table}

\subsection*{C. The relationship between frame length and recognition results}
Table~\ref{tabC} shows the recognition results obtained using different frame lengths. The input feature for all experiments is the STFT spectrogram, and the model backbone for all experiments is ResNet with multi-head attention. No regularization or augmentation technique is applied. The results indicate that setting the frame length to 50ms is the optimal choice for all three databases.

\begin{table}[ht]
\normalsize
    \caption{\label{tabC} The relationship between frame length and recognition results.\\}
    \centering
	\scalebox{0.8}{
	\begin{tabular}{llccc}
        \hline
	Frame length/Frame shift  &Shipsear & DTIL&DeepShip\\
	\hline
	30ms/15ms&  72.78 & 95.10& 73.48 \\
        50ms/25ms&  \textbf{75.24} & \textbf{95.93}& \textbf{74.68} \\
        100ms/50ms& 72.78& 95.77 & 74.25 \\
	150ms/75ms& 74.56 & 95.31 & 71.97  \\

        \hline
	\end{tabular}}
\end{table}

\subsection*{D. Network Parameters and Training Time Cost}
In this work, we implement several baselines that are widely applied in underwater acoustic target recognition tasks, including random forest, support vector machines, fully convolution networks, Mobilenet-v3 small~\citep{howard2019searching}, ResNet-18~\citep{he2016deep}, SE-ResNet-18~\citep{hu2018squeeze} and ResNet-18 with attention mechanism.

Two classic machine learning methods are implemented as weak baselines: random forest, which is an ensemble learning method that uses multiple decision trees to make predictions, and support vector machines, which are a type of classification model that navigates data by maximizing the margin between different classes. In addition, we implement several methods based on deep neural networks. Fully convolutional networks (FCN) with 1-d convolution employ three one-dimensional convolution layers with different convolution kernel sizes (8$\times$8, 5$\times$5, 3$\times$3). MobileNet-v3 small employs techniques such as depth-wise separable convolutions, which replace a standard convolution with a depth-wise convolution and a point-wise convolution. ResNet consists of a 7$\times$7 convolutional layer, a 3$\times$3 max pooling layer, four basic blocks (see Figure 2), an adaptive average pooling layer, and a fully-connected layer. SE-ResNet is a ResNet-style model that includes a squeeze-and-excitation (SE) mechanism designed to improve the quality of the output feature maps. ResNet with multi-head attention is detailedly introduced in Section 2.2. The number of parameters and training time cost on Shipsear is illustrated in Table~\ref{tabD}.

\begin{table*}[ht]
\normalsize
    \caption{\label{tabD} The number of parameters and training time cost on Shipsear.}
    \centering
	\scalebox{0.85}{
	\begin{tabular}{llll}
        \hline
	Model&Methods&Para num(MB)& Time cost(mins)\\
	\hline
	FCN(1-d convolution)&-&0.545 &3.9 \\
        MobileNet-v3 small&-&2.425 &9.5 \\
        ResNet-18&-&10.657 & 26.4\\
        SE-ResNet-18&-&7.234 & 24.8 \\
        ResNet-18+linear attention&-&11.149& 27.0 \\
        ResNet-18+multi-head attention&-&11.841 & 27.6\\
         &+noise aug& 11.841& 48.0\\
         &+GANs aug& 11.841& 97.8\\
         &+smooth reg& 11.841& 47.6 \\
         &+LMR& 11.841&28.0 \\
         &+smooth reg, LMR&11.841 &49.5 \\
        \hline
	\end{tabular}}
\end{table*}

\subsection*{E. Class Activation Mapping for four spectrogram-based features}
\begin{figure*}[h]
    \centering
    \subfigure[The STFT spectrogram and the corresponding class activation mapping heat map.]{
        \begin{minipage}[b]{1\textwidth}
        \includegraphics[width=0.75\linewidth]{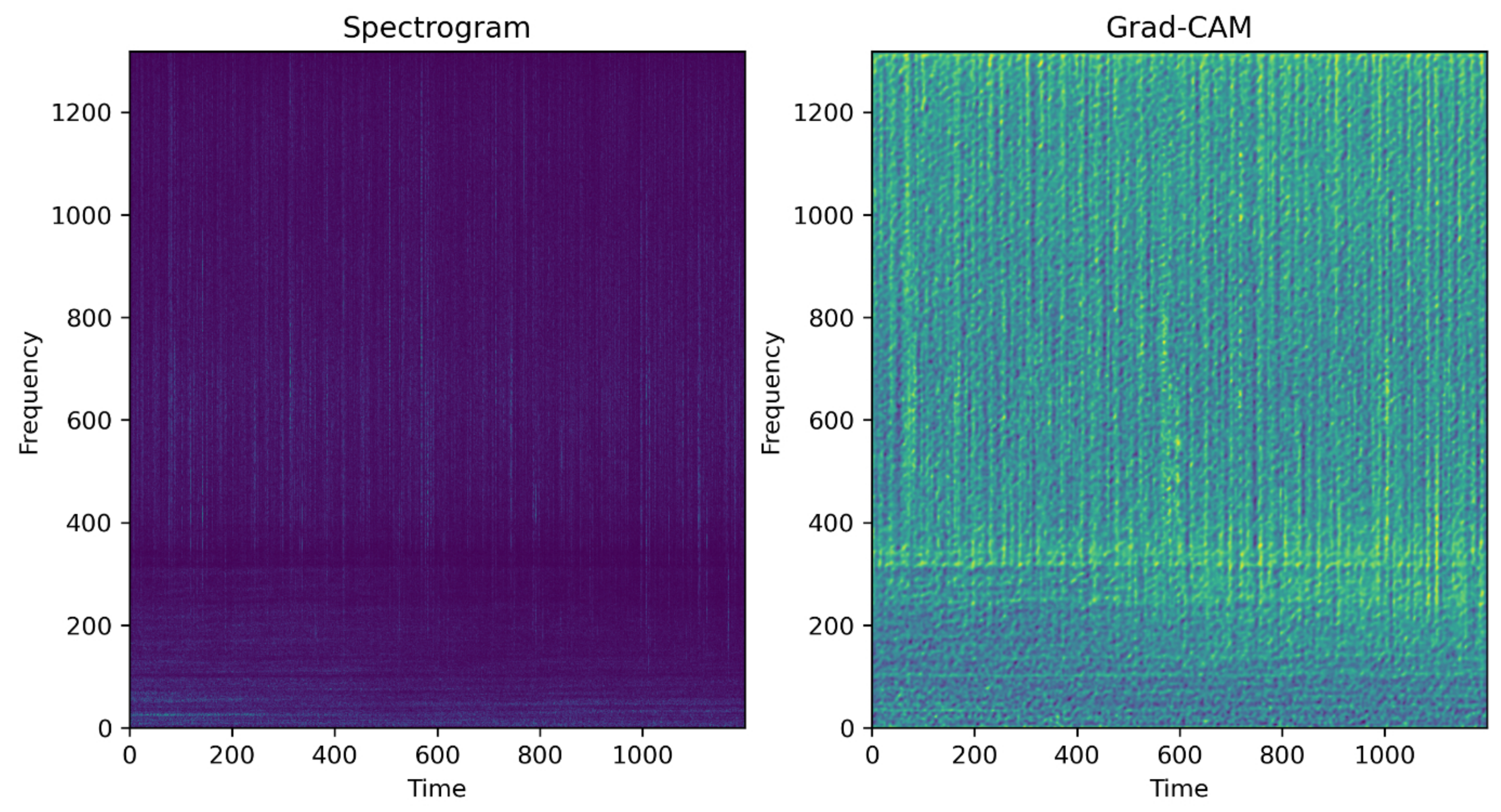}
        \centering
        \end{minipage}}

    \subfigure[The Mel spectrogram and the corresponding class activation mapping heat map.]{
        \begin{minipage}[b]{1\textwidth}
        \includegraphics[width=0.75\linewidth]{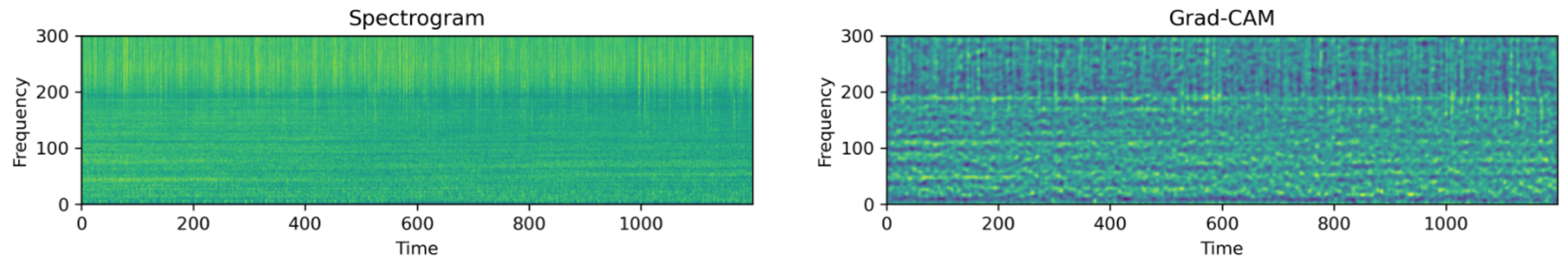}
        \centering
        \end{minipage}}

    \subfigure[The Bark spectrogram and the corresponding class activation mapping heat map.]{
        \begin{minipage}[b]{1\textwidth}
        \includegraphics[width=0.75\linewidth]{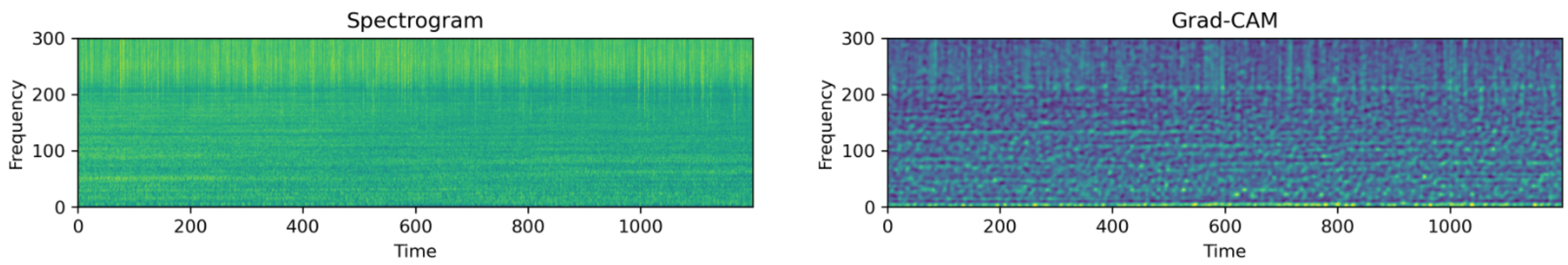}
        \centering
        \end{minipage}}

    \subfigure[The CQT spectrogram and the corresponding class activation mapping heat map.]{
        \begin{minipage}[b]{1\textwidth}
        \includegraphics[width=0.75\linewidth]{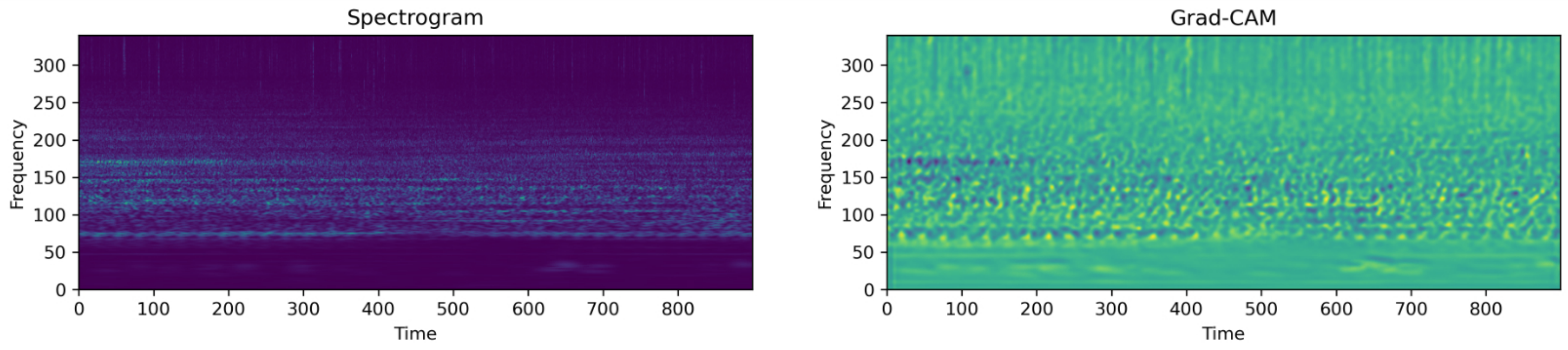}
        \centering
        \end{minipage}}
        
    \caption{The spectrograms and their corresponding CAM heat maps across four features. We take the 30-second segment of a ro-ro ship in Shipsear as the example.}
    \label{fig8}
    \vspace{-2px}
\end{figure*}

This work is based on four spectrogram-based acoustic features. In Appendix E, we use a 30-second segment (belonging to ro-ro ship) in Shipsear to exemplify the STFT, Mel, Bark, CQT spectrograms, and their corresponding class activation mappings, as shown in Figure~\ref{fig8}. We utilize CAMs to highlight important regions in spectrograms, which is conducive to exploring which parts of the features play a crucial role in recognition.

\newpage

\printcredits

\bibliographystyle{cas-model2-names}

\bibliography{cas-refs-advlmr}





\end{document}